\documentstyle[preprint,epsf,aps]{revtex}
\begin{document}
\title{Phase diagram of a generalized Hubbard model applied to orbital 
order in manganites}
\author{Qingshan Yuan$^{1,3}$, Takashi Yamamoto$^1$ and Peter Thalmeier$^2$}
\address{$^1$Max-Planck-Institut f\"{u}r Physik komplexer Systeme, 
N\"{o}thnitzer Str.38, 01187 Dresden, Germany\\
$^2$Max-Planck-Institut f\"{u}r Chemische Physik fester Stoffe, 
N\"{o}thnitzer Str.38, 01187 Dresden, Germany\\
$^3$Pohl Institute of Solid State Physics, Tongji University, 
Shanghai 200092, P.R.China}
\maketitle
\begin{abstract}
The magnetic phase diagram of a
two-dimensional generalized Hubbard model proposed for manganites is
studied within Hartree-Fock approximation. In this model the hopping matrix 
includes anisotropic
diagonal hopping matrix elements as well as off-diagonal elements. The 
antiferromagnetic (AF), ferromagnetic (F),
canted (C) and paramagnetic (P) states are included in the analysis as 
possible phases. 
It is found that away from half-filling only the 
canted and F states may exist and AF and P states which are
possible for the usual Hubbard model do not appear. This is because the 
F order has already developed for on-site repulsion $U=0$ due to the
hopping matrix of the generalized model. When applied for manganites the 
orbital degree is described
by a pseudospin. Thus our ``magnetic'' phase diagram obtained physically 
describes how orbital order changes with $U$ and with doping for manganites. 
Part of our results are consistent with other numerical calculations and 
some experiments. 

\end{abstract}


The three-dimensional (3D) cubic manganites R$_{1-x}$A$_x$MnO$_3$ (R is a
rare earth element such as La or Nd, A is a divalent alkali such as Sr or Ca)
and the layered ones La$_{n-nx}$Sr$_{1+nx}$Mn$_n$O$_{3n+1}$
($n=1,2$) have attracted intensive interest due to their rich material properties
\cite{Coey}. Most prominently they show a colossal
magnetoresistance which may be qualitatively understood on the basis of double
exchange (DE) model or the more general ferromagnetic Kondo lattice model. 
However, many recent experimental findings such as the doping dependence of charge, orbital, spin ordered phases \cite{Ramirez} suggest that additional 
physical mechanisms beyond the DE model should be involved. 
The possible ingredients are: $e_g$ 
orbital degeneracy \cite{Shiba,Takahashi,Kilian}, electron correlation
\cite{Ishihara1} as well 
as electron-lattice interaction\cite{Yuan}, e.g., by Jahn-Teller coupling \cite{Millis}. 
Though there exist some works to study the complex interplay of spin, orbital 
and lattice degrees of freedom $\cite{Mizokawa,Koshibae,Maezono}$, it is still
very helpful to clarify 
the role of each individual mechanism separately with the assumption that 
other degrees of freedom are frozen out. Actually, experiments showed that the 
spins in many 3D manganites R$_{1-x}$A$_x$MnO$_3$ are 
ferromagnetically ordered or A-type 
antiferromagnetic ordered (ferromagnetic layers coupled 
antiferromagnetically) in a large region of the doped phases, and A-type 
antiferromagnetic ordered in the undoped phase
\cite{Urushibara,Ramirez}. This means that a perfect spin-polarized 
two-dimensional (2D) plane is always retained over large region of doping. 
Therefore, the spin degree of freedom may be discarded and the 
following effective 2D Hamiltonian with only orbital degree of freedom
is proposed \cite{Motome,Nakano}:
\begin{eqnarray}
{\cal H} & = & -\sum_{<ij>}\sum_{\sigma \sigma'} t_{ij}^{\sigma \sigma'}
c_{i\sigma}^{\dagger}
c_{j\sigma'}+ U\sum_{i}(n_{i\uparrow}-1/2)(n_{i\downarrow}-1/2) \ ,\label{H}
\end{eqnarray}
where $<ij>$ indicates summation over nearest-neighbor sites, 
$t_{ij}^{\sigma \sigma'}$ denotes the hopping integral and $U$ is the effective
inter-orbital Coulomb interaction. The two orbitals $|x^2-y^2\rangle$ 
and $|3z^2-r^2\rangle$ have been assumed as pesudospin $\uparrow$ and $\downarrow$, 
respectively. The hopping matrix is explicitly 
given by $t_{ij}^{\uparrow\uparrow}=t_1=3t/4,\ t_{ij}^{\downarrow\downarrow}=t_2=t/4,\ 
t_{ij}^{\uparrow\downarrow}=t_{ij}^{\downarrow\uparrow}=-(+)t_3=-(+)\sqrt{3}t/4$
along the $x(y)$-direction 
\cite{Kugel}. The most important feature
here is the orbital dependence of the integral $t_{ij}^{\sigma \sigma'}$, 
which distinguishes the present model from the usual Hubbard model where we 
have $t_{ij}^{\sigma \sigma'}=t\delta_{\sigma \sigma'}$. 

Although the above 2D model ignores some important effects of three
dimensionality as we discuss in the later, it may still contain important
physics for 3D cubic manganites. For this the above model or its strong
coupling version, the so-called orbital $t$-$J$ model has been used to
study the unusual optical conductivity observed in doped LaMnO$_3$
$\cite{Nakano,Horsch}$. At the same time it should be also noted that 
the above model may be of direct relevance to layered manganites, 
which consist of stacking of single or double MnO$_2$ layers. 
Thus a full understanding of the properties of the
above model is necessary. While its optical conductivity was studied by
others, we discuss a different aspect here. It is natural to ask in which 
properties the present model deviates from the usual Hubbard model.
For example, as a basic problem, what are the differences between their 
magnetic (it actually means ``orbital'' for the real manganites) phase 
diagrams? This will be the main topic addressed in this work. Though 
the orbital ordering for manganites based on the current 
model has been numerically studied at some points \cite{Motome,Horsch}, 
the full phase diagram has not been discussed. So the purpose of this paper 
is twofold: On one hand we want to know how its magnetic phase diagram
is modified by the generalized hopping matrix, in analogy to the recent
investigation of the influence of next-nearest-neighbor hopping 
$\cite{Hlubina}$ on the phase diagram. On the other hand, as 
a real model for manganites, our ``orbital'' phase diagram obtained below
is certainly useful for the nontrivial question of how the orbital order is 
changed upon doping.

\begin{figure}[h]
\epsfxsize=8cm
\epsfysize=2.5cm
\centerline{\epsffile{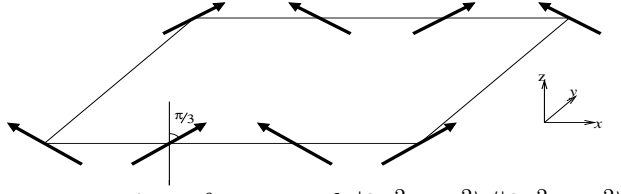}}
\caption{Pseudospin representation of staggered 
$|3x^2-r^2\rangle/|3y^2-r^2\rangle$ 
orbital ordering in $xy$ plane under the basis: $|x^2-y^2\rangle$ ($\uparrow$) and 
$|3z^2-r^2\rangle$ ($\downarrow$).}
\label{fig_spin}
\end{figure}

In the following we use the Hartree-Fock (HF) approximation to study 
the magnetic phases of the Hamiltonian (\ref{H}). Although the HF
approximation may overestimate the appearance of the ordered states due to
neglect of quantum fluctuations, this simple approximation may
still give many useful information. So it should be kept in mind that the 
quantitative aspects of the results given here should not be taken 
literally but that these results provide a basis for further more elaborated 
theoretical studies. In this respect the so called slave-boson mean field
theory is an improved method, but requires much more numerical efforts
\cite{Denteneer}.
Besides the usual ferromagnetic (F), 
antiferromagnetic (AF) ordered and paramagnetic (P) states considered 
here the canted ordered (C) state
is also taken into account as a possible phase. Although it has rarely 
been considered in previous studies of the Hubbard model, we include it here 
motivated by the fact that experiments on LaMnO$_3$ \cite{Murukami} have confirmed 
the type of $|3x^2-r^2\rangle/|3y^2-r^2\rangle$ orbital ordering which corresponds to 
a canted spin order in the pesudo-spin representation \cite{Remark}, as
illustrated in Fig. \ref{fig_spin}. Other complicated spin ordered states 
which have been studied for the usual Hubbard model such as spiral phase 
\cite{Dzierzawa} etc. will not be considered.

In HF approximation the Hamiltonian (\ref{H}) can be rewritten as 
\cite{Fradkin} (irrelevant constants are ignored):
\begin{eqnarray}
{\cal H}_{\rm HF} & = & -\sum_{<ij>}\sum_{\sigma \sigma'} t_{ij}^{\sigma 
\sigma'}c_{i\sigma}^{\dagger}
c_{j\sigma'} -{2\over 3}U\sum_{i}(2\vec{S}_i \cdot \langle\vec{S}_i\rangle-
\langle\vec{S}_i\rangle\cdot \langle\vec{S}_i\rangle) \ ,\label{H_HF}
\end{eqnarray}
where $\vec{S}_i={1\over 2}\sum_{\sigma \sigma'}c_{i\sigma}^{\dagger}
\vec{\sigma}_{\sigma \sigma'}c_{i\sigma'}$ and $\vec{\sigma}$ are Pauli 
matrices. We restrict to solutions with uniform electron density and canted
order of the magnetic moments characterized by 
\begin{eqnarray}
\langle\vec{S}_i\rangle=\left\{ \begin{array}{ll}
m(\sin \theta,\ 0,\ \cos \theta) & \ \ i\in {\rm A}\ ,\\
m(-\sin \theta,\ 0,\ \cos \theta) &\ \  i\in {\rm B}\ .
\end{array}\right. \label{S}
\end{eqnarray}
Here we have assumed a bipartite lattice with sublattices A and B.
For later discussion it is indispensable to understand the nature
of all possible ordered states characterized by the parameter 
$\theta$ (with $m$ describing the size of the
order parameter), as well as their corresponding orbital 
ordering \cite{Remark}. 
\begin{enumerate}
\item $\theta=0\ (\pi)$: F order with spin $\uparrow$ ($\downarrow$) or 
   homogenous $|x^2-y^2\rangle$ ($|3z^2-r^2\rangle$) orbital ordering.
\item $\theta=\pi/2$: AF order or staggered 
($|x^2-y^2\rangle+|3z^2-r^2\rangle$)
/($|x^2-y^2\rangle-|3z^2-r^2\rangle$) orbital ordering.
\item other $\theta$ values: canted-F or AF order. For example, 
  $\theta=\pi/3\ (2\pi/3)$ corresponds to staggered 
 $|3x^2-r^2\rangle/|3y^2-r^2\rangle$ ($|y^2-z^2\rangle/|x^2-z^2\rangle$) 
 orbital ordering.
\end{enumerate}
In addition we note that due to broken SU(2) symmetry by the hopping matrix in
Eqs. (1,2) the direction of spin polarization in the ordered states is no longer
rotationally invariant, contrary to the case for the usual Hubbard model. For
example, the F phases with spin $\uparrow$ and $\downarrow$ are not equivalent 
now. We also remark that the spins are lying within the plane in the AF 
ordered state described by Eq. (\ref{S}).

With new fermion operators $a$ and $b$ corresponding to sublattice A and B, 
respectively, we may write the Hamiltonian (\ref{H_HF}) in the momentum
space as  ${\cal H}_{\rm HF}=2NUm^2/3 +(a_{k\uparrow}^{\dagger}\ b_{k\uparrow}
^{\dagger}\ a_{k\downarrow}^{\dagger}\ b_{k\downarrow}^{\dagger}){\cal H}^{\rm M}
(a_{k\uparrow}\ b_{k\uparrow}\ a_{k\downarrow}\ b_{k\downarrow})^{\rm T}$ with
${\cal H}^{\rm M}$ denoting a $4\times 4$ Hermitian matrix.
Here $N$ is the total number of lattice sites and the reduced Brillouin 
zone is 
$|k_x|+|k_y|\le \pi$ or $-\pi/2<k_1,\ k_2\le \pi/2$
($k_{x,y}=\pm k_1+k_2$). In principle ${\cal H}^{\rm M}$ leads to four 
energy bands 
$\varepsilon_k^{\alpha}$
($\alpha=1,2,3,4$). The ground state energy is given by 
$E_{\rm GS}=\sum_{k,\alpha}\varepsilon_k^{\alpha}+2NUm^2/3$, where the 
first contribution depends on the band filling
$n=N_e/N$ where $N_e$ is the total electron number. Next we calculate 
numerically the values $m$, $\theta$ by minimizing the $E_{\rm GS}$ for given
repulsion $U$ and band filling $n$ or doping $\delta$ ($=1-n$). For this purpose we have used a mesh of 
$10^2\times 10^2$ $k$ points in the Brillouin zone. In the following discussion of results $t$ is set as energy unit.

\begin{figure}[h]
\epsfxsize=8cm
\epsfysize=6cm
\centerline{\epsffile{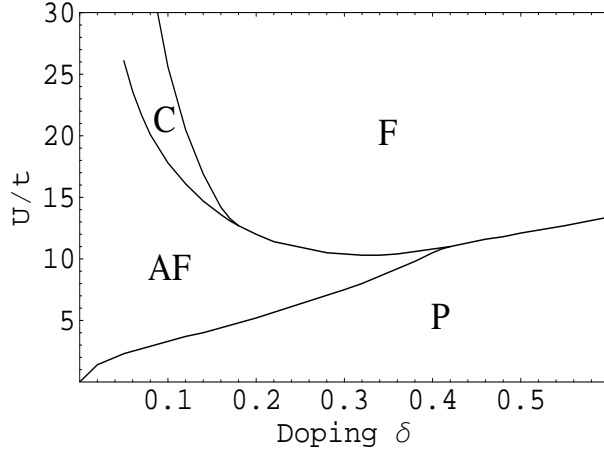}}
\caption{Magnetic phase diagram for the usual Hubbard model with inclusion of 
ferromagnetic (F), antiferromagnetic (AF), canted (C) ordered and paramagnetic
(P) states.}
\label{fig_PH1}
\end{figure}

\begin{figure}[h]
\epsfxsize=8cm
\epsfysize=6cm
\centerline{\epsffile{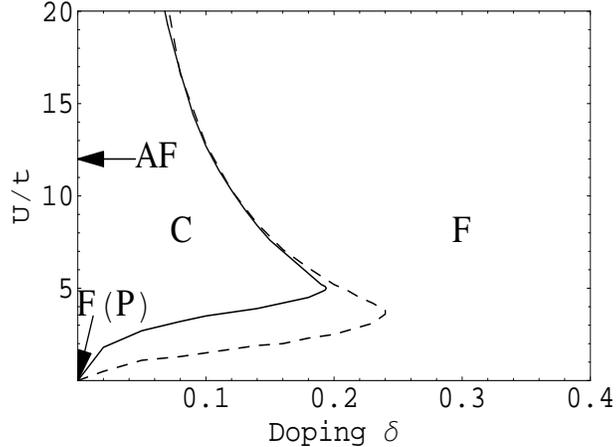}}
\caption{Magnetic phase diagrams. The solid (dashed) line are for the generalized Hubbard model with $t_1=3/4,t_2=1/4,t_3=\sqrt{3}/4$ 
($t_1=3/4,t_2=1/4,t_3=0$) respectively. F(P) denote the stable phase at U=$\delta$=0 for both cases.}
\label{fig_PH2}
\end{figure}

With values of $m$, $\theta$ obtained from the minimization of E$_{GS}$ 
we may plot the magnetic phase diagrams 
in the plane ($\delta, U$), as shown by Figs. \ref{fig_PH1} and \ref{fig_PH2}. 
Although there are a lot of studies for the usual Hubbard model (i.e.,
$t_1=t_2=1,t_3=0$) \cite{Denteneer},
we give its phase diagram again in Fig. \ref{fig_PH1} 
since the canted ordered state is added here. We also want to compare it
with the result for the current
generalized Hubbard model whose phase diagram is displayed in
Fig. \ref{fig_PH2} (solid line). In the present model, the hopping matrix involves anisotropic diagonal elements and
off-diagonal ones. To see more clearly how the phase diagram is influenced by
them, we plot it for a simplified model with a hopping matrix which
only includes anisotropic diagonal elements (i.e.,
$t_1=3/4,t_2=1/4,t_3=0$); it is denoted by a dashed line in Fig. \ref{fig_PH2}. 

>From Figs. \ref{fig_PH1} and \ref{fig_PH2}, a number of 
differences can be seen. 
Firstly, the phases for the usual Hubbard model include AF, F, P
states as well as a narrow region with a canted phase. It appears in a
crossover region from AF to F phase, which is understandable. 
The phase diagram is then qualitatively modified by introduction of
anisotropic diagonal elements where only the canted and F ordered phases exist
and the original AF and P phases are excluded away from half-filling 
(i.e., $\delta >0$). Furthermore when the off-diagonal
elements are introduced the phase diagram is mainly quantitatively modified, 
i.e., the stability region of the F ordered phase is enlarged. We point out 
that the F phase in 
Fig. \ref{fig_PH2} always means that spins align in the up direction 
($\uparrow$) due to a 
larger hopping matrix element $t_{ij}^{\uparrow\uparrow}$ (i.e., $t_1>t_2$).

\begin{figure}
\epsfxsize=8cm
\epsfysize=6cm
\centerline{\epsffile{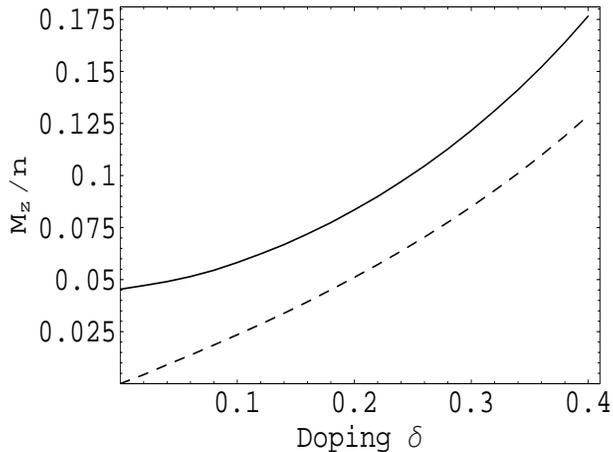}}
\caption{The $z$-axis magnetization for the generalized Hubbard model
(solid line: $t_3=\sqrt{3}/4$, dashed line: $t_3=0$) for on-site repulsion 
$U=0$.} 
\label{fig_Mz}
\end{figure}

The above results may be understood if we consider the property of the
kinetic term alone, i.e., assume that $U=0$. Then the Hamiltonian
(\ref{H}) may be exactly diagonalized with two bands:
\begin{eqnarray}
\epsilon_k^{\pm} & = & -(t_1+t_2)(\cos k_x +\cos k_y)\pm
\sqrt{(t_1-t_2)^2(\cos k_x +\cos k_y)^2+4t_3^2(-\cos k_x +\cos k_y)^2}\ .
\end{eqnarray}
>From $\epsilon_k^{\pm}$ the $z$-axis magnetization
$M_z=\langle c_{i\uparrow}^{\dagger}c_{i\uparrow}-c_{i\downarrow}^{\dagger}
c_{i\downarrow}\rangle/2$ may be 
calculated, as shown in Fig. \ref{fig_Mz} where the curves $M_z/n$ vs $\delta$
are given. For the usual Hubbard model $M_z$ is always zero,
corresponding to P state for the abscissa in Fig. \ref{fig_PH1}.
 However,  
a finite magnetization (with spin $\uparrow$) is always exhibited for the
present model as shown by the solid line in Fig. \ref{fig_Mz}; also for the 
simplified model (zero off-diagonal hopping)  shown by the dashed line, except at the point $\delta=0$.  
Moreover, the $M_z$ value for the current model is always bigger than that
for the simplified one, which means that the F order of the former is stronger. 
Because of the underlying F order already in the $U=0$ case only the canted 
(with nonzero F component
involved) and F ordered phases may exist in finite $U$ region away from
half-filling. It is also easily understood why the F phase region is enlarged
after inclusion of the off-diagonal hopping matrix elements.

We consider the case of half-filling (i.e., $\delta =0$). It has been shown
that the usual Hubbard model has an AF ground state for any finite $U$, 
corresponding to the ordinate in the phase diagram of
Fig. \ref{fig_PH1}. We speculate that this conclusion might also be true 
for our present
generalized Hubbard model. Although the present ground state is F ordered at 
$U=0$ as given in Fig. \ref{fig_Mz}, it is found to have AF order for $U>1.2$.
This was concluded from a calculation employing a mesh of $10^3\times 10^3$ $k$
points. Whether or not there
exists a crossover between F and AF for finite $U$ is not yet certain.

\begin{figure}
\epsfxsize=8cm
\epsfysize=6cm
\centerline{\epsffile{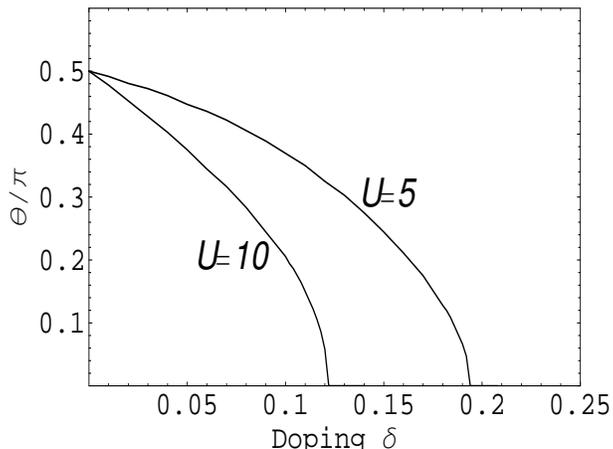}}
\caption{The variation of parameter $\theta$ (in units of $\pi$) 
characterizing the orbital ordering with doping $\delta$ for fixed
$U$ for the generalized Hubbard model (\ref{H}).}
\label{fig_Th}
\end{figure}

Our ``magnetic'' phases obtained from Hamiltonian (\ref{H}) actually 
correspond to the
orbital orderings relevant for the manganites as mentioned before, therefore we
now discuss them in detail. To
see clearly how the orbital order changes with doping $\delta$
we have plotted in Fig. \ref{fig_Th} the relation of parameter $\theta$ with
$\delta$ for fixed $U$. It is found $\theta$ decreases monotonically from
$\pi/2$ to $0$ with increase of $\delta$ indicating the process AF$\rightarrow$ 
C$\rightarrow$ F order for the pseudospins as shown in Fig. \ref{fig_PH2}. 
Let us take $U=10$ as an example in the following. (The real value
for it in manganites is about $4\sim 10$\cite{Takahashi}). In the undoped case, $\theta=\pi/2$ means a staggered 
($|x^2-y^2\rangle+|3z^2-r^2\rangle$)/($|x^2-y^2\rangle-|3z^2-r^2\rangle$)
orbital ordering, which is consistent with the result by Motome and Imada 
from quantum Monte Carlo calculation\cite{Motome}, but slightly
different from the experimentally observed $|3x^2-r^2\rangle/|3y^2-r^2\rangle$
\cite{Murukami}. This may be due to some other physical mechanism, e.g, 
a Jahn-Teller effect which is present in real materials but not included here 
\cite{Motome}. On the other hand, it is found that the 
$|3x^2-r^2\rangle/|3y^2-r^2\rangle$ type orbital
ordering may be also exhibited from our pure orbital model at very small doping
$\delta \simeq 0.06$. With increase of doping the orbital
order changes continuously until at some critical value $\delta_c \simeq 0.12$
which is still small it becomes $|x^2-y^2\rangle$ type. This behaviour for
small $\delta_c$ has also been noted by Horsch {\it et al.} based on the
orbital $t$-$J$
model (i.e., large $U$ version of ours) with the exact diagonalization method
\cite{Horsch}. Interestingly it is also seen that the value $\delta_c$ is not
a monotonic function of $U$, which reaches a maximum $\sim 0.2$ at about
$U=5$. Recently Akimoto {\it et al.} found an A-type 
antiferromagnetic metallic ground state for La$_{1-x}$Sr$_x$MnO$_3$ at
doping $x>0.5$, and they proposed that $|x^2-y^2\rangle$ orbitals should be
occupied in this state\cite{Akimoto}. This is consistent with our result
derived from the planar model (\ref{H}). In addition, we also noticed the
experimental findings that many cubic manganites are 3D ferromagnets in the
intermediate doping regime nearby $\delta=0.3$ \cite{Urushibara,Ramirez},
where the hopping along the $z$-axis is allowed by the DE mechanism.
Thus it is instructive to discuss the effect of three dimensionality 
neglected in our model in this doping region, see also Ref. \cite{Nakano}. 
For a 3D model, the hopping along the $z$-axis is orbitally 
highly anisotropic; only the hopping between two $|3z^2-r^2\rangle$
orbitals is permitted. So the
2D ferro-orbital order with $|x^2-y^2\rangle$ type polarization, as found
above, may be suppressed in order to gain the kinetic energy along the
$z$-axis. Such a depolarization may be simulated by introducing a 
negative chemical potential $\mu$ on the $|3z^2-r^2\rangle$
orbital \cite{Nakano}. It is found that, with reasonable $\mu$ valid for 3D
manganites the disordered orbital liquid state may appear at some doping
nearby $\delta=0.3$. This orbital liquid was first discussed by Ishihara
{\it et al.} \cite{Ishihara2}, which may be enhanced by the quantum 
fluctuations beyond HF approximation \cite{Comment}. On the other hand, 
for the layered manganites with single or double layers of MnO$_2$
where the inter-layer or bilayer hopping
(or correspondingly $\mu$) is assumed very small, the orbital liquid 
state may be excluded and our ferro-orbital ordered state is still expected
to exist in the doping region mentioned above.  

In the above discussion we have shown the possible orbital
ordering as function of doping $\delta$ and repulsion $U$. The results
are partly consistent with previous numerical calculations, and also with 
some experimental results on manganites although only orbital degree of 
freedom is retained in the present model. For a full understanding of various 
phases observed in manganites experimentally, one needs  
to consider the spin and orbital orderings simultaneously 
\cite{Koshibae,Maezono}, as well as include lattice degree of freedom in some
cases \cite{Motome}. However, our orbitally 
ordered structures obtained above are very instructive to understand the 
role of orbital degree of
freedom itself, i.e., how the interplay of the orbital kinetics (with
attention to its peculiar matrix character) and orbital
correlation induces the orbital ordering. Moreover, they may be also possibly 
observed under some conditions, e.g., at large external magnetic fields.     

In summary, within HF approximation we have studied the magnetic phase diagram
of a generalized Hubbard model proposed for manganites where the hopping
matrix includes anisotropic
diagonal elements and off-diagonal elements. The AF, F, canted ordered 
and P states
are considered as possible phases. It is found that in the doped case only the 
canted and F states may exist, while the AF and P phases of the usual
Hubbard model are not favored. This is because the F order has already been 
established at $U=0$ due to the peculiar hopping matrix. When applied for manganites, the spin index of this model actually represents the orbital degree of
freedom. Thus our obtained magnetic phase diagram is able to describe how
the orbital order changes upon doping. 
  
\bigskip

Two of authors (Q. Yuan and T. Yamamoto) would like to thank the support of
Visitor Program of MPI-PKS, Dresden. Q. Yuan also acknowledges the part support by
Chinese NSF. 

\bigskip
{\it Note Added in Proof:} 
Recently the scenario for phase separation has been extensively
studied for various models in the manganites \cite{Note}. Such a tendency
of separation into orbitally ferro- and antiferro- ordered regions
may also exist in our case of the canted phase in Fig. 3. This problem
deserves further detailed investigation.


\begin{references}
\bibitem{Remark} An arbitrary orbital state may be defined by the linear 
  combination of two orbitals as $\cos \phi |3z^2-r^2\rangle+
  \sin \phi |x^2-y^2\rangle$, or equivalently 
  represented as a pseudospin vector $(\sin \phi',0,\cos \phi')$ with relation
  $\phi'=\pi-2\phi$. For example,
  the states $|3x^2-r^2\rangle,\ |3y^2-r^2\rangle,\ |y^2-z^2\rangle$ and 
  $|x^2-z^2\rangle$ correspond
  to $\phi=2\pi/3,\ -2\pi/3,\ -5\pi/6$ and $5\pi/6$ respectively.
\end{references}
\end{document}